\documentclass{ifacconf}

\usepackage{graphicx}      
\usepackage{natbib}        
\usepackage{amsfonts}
\usepackage{amsmath}
\usepackage{wrapfig}
\theoremstyle{definition}
\newtheorem{asm}{Assumption}
\usepackage{graphicx}
\usepackage{caption}
\usepackage{subcaption}

\usepackage{bm} 
\usepackage{amssymb}
\usepackage[subrefformat=parens]{subcaption}
\usepackage{comment}
\usepackage{xcolor}

\makeatletter
\newcommand{\IFACcopyrightbox}{%
  \fbox{%
    \parbox{1\textwidth}{\footnotesize\centering
      ©\the\year\ the authors.\enspace
      This work has been accepted to IFAC for publication under a
      Creative Commons Licence CC-BY-NC-ND.}}}

\def\ps@copyright{%
  \let\@mkboth\@gobbletwo          
  \def\@oddhead{\@logohead}%
  \let\@evenhead\@oddhead%
  \def\@oddfoot{\hfill\IFACcopyrightbox\hfill}%
  \let\@evenfoot\@oddfoot}
\makeatother
\begin{document}
\begin{frontmatter}

\title{Risk-Aware Trajectory Optimization and Control for an Underwater Suspended Robotic System\thanksref{footnoteinfo}} 

\thanks[footnoteinfo]{This work was supported by the DAAD program Konrad Zuse Schools of Excellence in Artificial Intelligence, sponsored by the Federal Ministry of Education and Research, and by the European Union’s Horizon Europe innovation action program under grant agreement No. 101093822,
”SeaClear2.0”.}

\author[1st]{Yuki Origane\thanksref{footnotecontrib}}
\author[2nd]{Nicolas Hoischen\thanksref{footnotecontrib}}
\author[2nd]{Tzu-Yuan Huang}
\author[1st]{Daisuke Kurabayashi}
\author[2nd]{Stefan~Sosnowski}
\author[2nd]{Sandra Hirche}

\thanks[footnotecontrib]{Equal Contribution}

\address[1st]{Institute of Science Tokyo, 
   2-12-1 Ookayama, Meguro-ku, Tokyo 152-8552, Japan,  (e-mail: \{origane,dkura\}@irs.sc.e.titech.ac.jp).}
\address[2nd]{ Chair of Information-oriented Control, Technical University of Munich, Germany, (e-mail: {\{nicolas.hoischen, tzu-yuan.huang, sosnowski, hirche\}@tum.de}).}

\begin{abstract}                

This paper focuses on the trajectory optimization of an underwater suspended robotic system comprising an uncrewed surface vessel (USV) and an uncrewed underwater vehicle (UUV) for autonomous litter collection. The key challenge lies in the significant uncertainty in drag and weight parameters introduced by the collected litter. We propose a dynamical model for the coupled UUV-USV system in the primary plane of motion and a risk-aware optimization approach incorporating parameter uncertainty and noise to ensure safe interactions with the environment. A stochastic optimization problem is solved using a conditional value-at-risk framework. Simulations demonstrate that our approach reduces collision risks and energy consumption, highlighting its reliability compared to existing control methods. 

\end{abstract}


\end{frontmatter}
\section{Introduction}
\begin{wrapfigure}{o}[0pt]{1.67in}  
  \centering
  \includegraphics[width=1.75in]{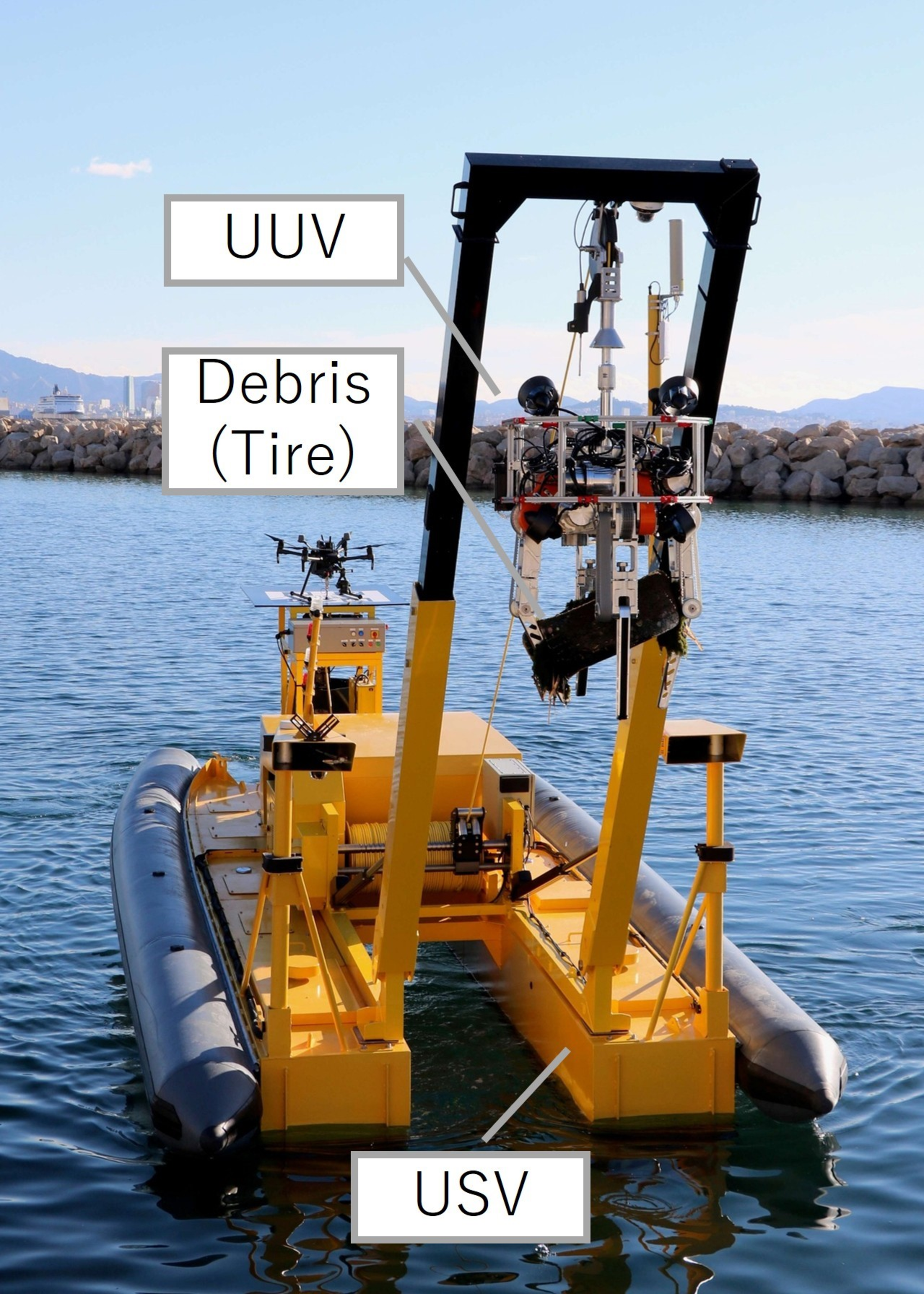}
  \caption{SeaClear USV-UUV System with the Seacat 2.0 from Subsea Tech.}
  \label{fig:problem_setting}
\end{wrapfigure}

Robotic solutions are increasingly vital for hazardous underwater environments, from offshore industry to marine civil engineering \citep{sivvcev2018underwater, zereik2018challenges}. In the growing field of ocean cleaning, for instance, the SeaClear project \citep{seaclear2} is developing a fully actuated grapple UUV to collect seafloor litter, a task once left to divers. 
The UUV is tethered to a winch on a USV to lift heavy debris (Figure \ref{fig:problem_setting}). The task of navigating the grapple UUV to a target position for litter collection and returning to the surface can be approached as a trajectory optimization problem for the coupled UUV-USV system. 
However, underwater navigation is challenging due to nonlinear UUV dynamics. The unknown mass and hydrodynamic coefficients of collected litter add further uncertainty, potentially causing oscillations that compromise stability. Moreover, obstacles and currents can deviate the UUV from its mission, demanding resilient and safe control to minimize the risk of harmful events.

Many guidance strategies separate planning \citep{petres2009trajectory, heo2013rrt} from tracking, addressing uncertainties late in the process \citep{beckers2022safe}. 
A more integrated approach, however, formulates a single optimization problem that incorporates prior knowledge of the robot's dynamics and environment. Enforcing safety within this problem leads to three main paradigms  \citep{Akella2024RiskAwareRT}. The \emph{worst-case} paradigm prioritizes safety by ensuring constraints hold under any disturbance, but is often overly conservative \citep{bansal2017hamilton,kuwata2005robust}. 
On the other hand, \emph{risk-neutral} methods optimize for expected outcomes but ignore low-probability, high-impact events critical for safety \citep{mesbah2016stochastic}. In contrast, \emph{risk-aware} strategies use tail-risk measures like Conditional Value-at-Risk (CVaR) to effectively distinguish between high- and low-variance distributions. This approach is appealing because it balances conservativeness with feasibility and can nuance the severity of unsafe events, such as differentiating a gentle seabed contact from a high-speed collision \citep{chow2015risk}. CVaR constraints quantify risk aversion by considering expected costs in the worst-case $\alpha \times 100\%$ scenarios. However, CVaR is often applied in MPC frameworks with linearized models \citep{hakobyan2019risk, sopasakis2019risk, dixit2023risk}, which is undesired for our highly nonlinear UUV-USV model, where nonlinearities help stabilize oscillations. A more suitable approach \citep{lew2023risk}, formulates general dynamics via stochastic differential equations (SDEs) and enforces CVaR constraints over entire trajectories rather than pointwise \citep{hakobyan2019risk}. The resulting stochastic OCP is then solved using Sample-Average Approximation (SAA) \citep{lew2024sample}.

The SeaClear2.0 project introduces a novel UUV-USV configuration (see Figure \ref{fig:problem_setting}), that has not been modeled before. In tethered UUV-USV systems, previous studies constrain the UUV's position via cable length \citep{morinaga2023trajectory}, model tether dynamics with infinitesimal segments  \citep{hong2020dynamics}, or mitigate tension using winches \citep{zhao2022offshore, zhao2021investigating}. These approaches typically assume low-tension, exploratory scenarios. In contrast, our work focuses on heavy debris transport, where the UUV is suspended by a constantly taut tether, allowing us to ignore cable bending while accounting for significant tension. Moreover, unlike fixed-platform models \citep{nikolas2022analysis}, our system features a mobile USV and an actively controlled UUV in a 2D pendulum-like configuration.

\textbf{Contributions }
First, we derive a model for the SeaClear UUV-USV system using Euler-Lagrange equations under a tensioned tether assumption. We then introduce a risk-aware trajectory optimization that extends existing open-loop, CVaR-based Sample Average Approximation (SAA) methods \citep{lew2024sample, lew2023risk} into a closed-loop scheme with a local feedback controller (RA-SAA+FB). This extension offers two key benefits: 1) it enables high-frequency online control via the local feedback while the main optimization runs at a lower frequency, and 2) it yields more compact, feasible trajectories by optimizing over a class of closed-loop controllers.  We perform Monte Carlo simulations to compare our method against competitive control strategies, including A*-based path planning with PID and Control Barrier Functions (CBFs), and Model Predictive Control Path Integral (MPPI) \citep{williams2018information}.


\textbf{Paper Organization}
Section \ref{sec:PbForm} presents the modeling of the USV-UUV system and the problem statement. Section \ref{sec:Method} outlines the proposed RAA-SAA approach with the feedback extension,  and Section \ref{sec:Sim} illustrates the results on our model against popular control approaches.

\section{PROBLEM FORMULATION}
\label{sec:PbForm}
This study focuses on navigating a UUV-USV system to a target location despite uncertainties (weight of the grasped litter, drag forces) and obstacles (position, size). The following section details the system's dynamical model and formulates this control objective.


\subsection{Modeling} \label{sec:modeling}
\begin{figure}[!t]
\centering
\includegraphics[width=2.0in]{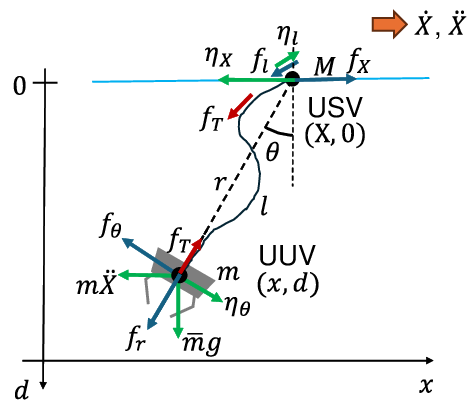}
\caption{Model of suspended UUV and USV}
\label{fig:model}
\end{figure}

To verify the fundamental performance of our method and facilitate comparison with existing approaches, we analyze a simplified model of the UUV-USV system with only three degrees of freedom. The model is illustrated in Figure~\ref{fig:model} and we can apply some model simplification due to the nonholonomic nature of the USV.
The USV and the UUV operate on a plane defined by the coordinates $(x,d)$.
Their positions at time $t$ are represented as $(X(t),0)$ and $(x(t),d(t))$ respectively.
Since we define the direction of $d$ as the depth, $d(t)>0$ means that the UUV is underwater at time $t$.
The UUV and the USV are connected by a tether.
The USV can adjust the tether length $l(t)>0$ using a winch.

To develop a tractable dynamic model of the tethered USV–UUV system, we make the following assumptions:

\begin{itemize}
\item The USV's propeller generates a strong propulsion force, allowing us to ignore the reaction force from the tether.
\item The tether is considered sufficiently thin and lightweight that its inertia, weight, and drag can be neglected for reasonably short tether lengths.
\item  The UUV is modeled as a point mass, and environmental effects such as ocean currents or wave-induced motion are disregarded.
\end{itemize}

Considering the second assumption, the dynamic model of the USV is given by
\begin{align}
M\ddot{X} = -\eta_X(\dot{X}) + f_X
\label{eq:eom_vessel}
\end{align}
Here, $M$ denotes the mass of the USV, $\eta_X({\dot X})$ represents the drag force from the USV, and $f_X$ is the propulsion force of the USV. 
We set up the model of the tether length $l$ with the mass of the winch $M_l$, the drag force  $\eta_l(\dot{l})$, the tension force $f_T$, and the winch force $f_l$.
\begin{align}
M_l \ddot{l} = -\eta_l(\dot{l}) + f_T + f_l\label{eq:eom_tether}
\end{align}
Next, to derive a dynamic model of the UUV, we define the kinetic energy $K$ and the potential energy $U$ as follows
\begin{align}
K = \frac{1}{2}m\left(\dot{x}^2+\dot{d}^2\right),\quad U = -\bar{m}gd
\end{align}
Here, $\bar{m}$ is the apparent UUV weight due to the difference between buoyancy and gravitational forces in water.
To simplify the tether force calculation, we transform the UUV's coordinates from $(x,d)$ to polar coordinates $(r,\theta)$, which represent the relative distance and angle of the UUV from the USV, respectively. The transformation is defined as:
\begin{align}
    x = X - r\sin\theta,\quad d = r\cos\theta, \label{eq:coordinate_transformation}
\end{align}
where $r > 0$ and $\theta \in (-\pi,\pi]$. Its time derivative is given by
\begin{align}
    \dot{x} = \dot{X} - \dot{r}\sin\theta - r\dot{\theta}\cos\theta,\quad \dot{d} = \dot{r}\cos\theta-r\dot{\theta}\sin\theta.\label{eq:velocity_transform}
\end{align}
Next, the Lagrangian $L = K - U$ yields
\begin{align}
    \begin{aligned}
    L =& \frac{1}{2}m(\dot{X}^2+\dot{r}^2+r^2\dot{\theta}^2)\\
    &-m\dot{X}\dot{r}\sin\theta-mr\dot{X}\dot{\theta}\cos\theta+\bar{m}gr\cos\theta.
    \end{aligned}\label{eq;lagrangian}
\end{align}

By applying the Euler-Lagrange equations, we derive the dynamics of the UUV:

\begin{align}
    \frac{d}{dt}\frac{\partial L}{\partial \dot{r}}-\frac{\partial L}{\partial {r}} 
    &= m\ddot{r}-m\ddot{X}\sin\theta-mr\dot{\theta}^2-\bar{m}g\cos\theta \notag \\
    &= -\eta_r-f_T+f_r,  \label{eq:eom_r}
    \end{align}
\begin{align}
    \frac{d}{dt}\frac{\partial L}{\partial \dot{\theta}}-\frac{\partial L}{\partial {\theta}}
    &= mr^2\ddot{\theta}-mr\ddot{X}\cos\theta+2mr\dot{r}\dot{\theta}+\bar{m}gr\sin\theta \notag 
    \\ &= r(-\eta_\theta+f_\theta),
\label{eq:eom_theta}
\end{align}
where the drag forces $\eta_r(\dot{r},\dot{X})$ and $\eta_\theta(\dot{\theta},\dot{X})$, resulting from the UUV's movement, along with the thruster forces in polar coordinate, $f_r$ and $f_\theta$, are explicitly defined.
Finally, we obtain the UUV-USV model
\begin{equation}
    \left\{
    \begin{aligned}
    M\ddot{X} &= -\eta_X(\dot{X})+f_X\\
    mr^2\ddot{\theta} &= m\ddot{X}r\cos\theta-2mr\dot{r}\dot{\theta}-\bar{m}gr\sin\theta\\
    &\quad +r(-\eta_\theta(\dot{\theta},\dot{X})+f_\theta)\\
    (m+M_l)\ddot{r} &= m\ddot{X}\sin\theta+mr\dot{\theta}^2+\bar{m}g\cos\theta\\
    &\quad -\eta_r(\dot{r},\dot{X})-\eta_l(\dot{r})+f_r + f_l\\
    \ddot{l} &= \ddot{r}
    \end{aligned}
    \right.\label{eq:hybrid_r_geq_l}
\end{equation}
To derive equation (\ref{eq:hybrid_r_geq_l}), we replace $f_T$ in (\ref{eq:eom_r}) with (\ref{eq:eom_tether}) and substitute the geometric constraint $\dot{l}=\dot{r}$ when the tether wire is under tension.

We assume that there is a first-order integrator delay between the force input ${\bm f} = (f_\theta, f_r, f_l, f_X)^\top$ generated by the actuators and the control signal ${\bm u} = (u_\theta, u_r, u_l, u_X)^\top$ sent from the computer to the actuators.
Let $T_\theta, T_r, T_l, T_X$ be the time delay constants. 

This relationship is expressed as follows
\begin{align} \label{eq:input_force_relation}
    \dot{\bm f}(t) = T^{-1}(-{\bm f}(t) + {\bm u}(t)),
\end{align}
where $T^{-1}\in\mathbb{R}^{4\times 4}$ is a diagonal matrix with diagonal components $1/T_\theta,1/T_r,1/T_l,1/T_X$.
We model the drag forces as follows
\begin{align}
    \eta_X(\dot{X}) &= c_X \dot{X},\quad \eta_l(\dot{l}) = c_l \dot{l},\\
    \eta_\theta(\dot{\theta},\dot{X}) &= c_\theta|\dot{X}\cos\theta+r\dot{\theta}|(\dot{X}\cos\theta+r\dot{\theta}),\\
    \eta_r(\dot{r},\dot{X}) &= c_r|\dot{X}\sin\theta+\dot{r}|(\dot{X}\sin\theta+\dot{r}),
\end{align}
where $c_X$, $c_l$, $c_r$, and $c_\theta$ are the viscosity coefficients of the drag forces.
Since $\eta_r$ and $\eta_\theta$ represent the drag forces acting on the UUV in the water, we model them as second-order expressions \citep{mathai2019dymamics}.

For the formulation of the OCP introduced later, we rewrite \eqref{eq:hybrid_r_geq_l} and \eqref{eq:input_force_relation} into the input-affine form
\begin{align}
    \dot{\bm x} = \bm{f}_{sys}({\bm x})+\bm{G}_{sys}({\bm x}){\bm u},\quad
    {\bm y} = \bm{h}({\bm x}).
    \label{eq:input_affine}
\end{align}
Here, we define the state vector as ${\bm q} = (\theta, r, l, X)^\top$ and ${\bm x} = ({\bm q}^\top, \dot{\bm q}^\top, {\bm f}^\top)^\top$. Additionally, we define ${\bm y} = \bm{h}({\bm x}) = (x, d, X)^\top$.
This transformation can always be performed under the assumption of this study that $r$ is positive.

\subsection{Control Objective}
By designing ${\bm u}(t)$, Our goal is to design a controller that satisfies
\begin{align}
    &\lim_{t\to t_f}\left(x(t),d(t),X(t)\right)\to(x_d,d_d,X_d)\\
    &{\rm s.t. }\quad \forall j,\forall t\in[0,t_f],\quad \|\left(x(t)-x_{\mathcal{O}_j}, d(t)-d_{\mathcal{O}_j}\right)\|^2\geq a_{\mathcal{O}_j},\nonumber
\end{align}
where $x_d,d_d,X_d$ are desired position of the UUV and USV and $t_f$ is a desired terminate time.
$x_d,d_d, X_d,t_f$ are given manually or by a task planning algorithm. $(x_{\mathcal{O}_j},d_{\mathcal{O}_j})$ represents the position of the convex obstacle $\mathcal{O}_j$, and $a_{\mathcal{O}_j}$ is size of circle which wraps the obstacle.
We assume the following when designing our controller

    \begin{asm}[Prior (Offline) Knowledge]  The mathematical model structure \eqref{eq:input_affine} is assumed to be known.
    Additionally, except for those mentioned in Assumption 2, the model parameters listed in Table \ref{tab:model_parameters} and the initial values $x(0), d(0), X(0)$ are assumed to be given.
    \end{asm}
    \begin{asm} [Uncertainty] 
    We consider $m,\bar{m},c_\theta,c_r$ as uncertain parameters, affected by the weight and shape of the collected litter.
    Additionally, the position $(x_{\mathcal{O}_j},d_{\mathcal{O}_j})$ and size of obstacles $a_{\mathcal{O}_j}$ are assumed to have uncertainty. These parameters are assumed to follow known probability distributions. Further, we assume that process noise following a Gaussian distribution with a known intensity is added to the system.
    \end{asm}
    \begin{asm} [Real Time (Online) Observations] The position, $x(t), d(t), X(t)$, and velocity, $\dot{x}(t), \dot{d}(t), \dot{X}(t)$, of the UUV-USV system are measurable.
    \end{asm}

In this study, we assume that position and velocity can be observed because these quantities can be relatively easily measured using accelerometers, pressure sensors, tether length and Doppler Velocity Logger information. On the other hand, the position of obstacles is considered to have low accuracy because it is measured using camera information in turbid water.
\subsection{Optimization Problem Formulation}
The uncertainty in the parameters and external disturbances affecting the UUV-USV during motion are captured by a stochastic model.
Therefore, we formulate our control objective as an Optimal Control Problem (OCP) subject to a Stochastic Differential Equation (SDE), which naturally models various sources of uncertainty and process noise.
Here we introduce the OCP and SDE for subsequent discussion, the mapping to our UUV-USV model is conducted in Section~\ref{sec:implementation}.
Let $(\Omega,\mathcal{G},\mathcal{F},\mathbb{P})$ be a filtered probability space, ${\bm W}$ be the $n$-dimensional standard Wiener process on $\Omega$.
${\bm x}_0\in\mathbb{R}^n$ is initial condition. We introduce an SDE
\begin{align*}
    d{\bm x}(t) &= \bm{b}({\bm x}(t),{\bm u}(t), {\bm \xi}(\omega))dt + \bm{\sigma}({\bm x}(t),{\bm u}(t), {\bm \xi}(\omega))d{\bm W}_t,\\
    {\bm x}(0) &= {\bm x}_0.
    \label{eq:SDE}
\tag{SDE}
\end{align*}
with solution $\bm{x}_u$. The functions \( \bm{b}:\mathbb{R}^n\times U\times\mathbb{R}^q\to\mathbb{R}^n \) and \( \bm{\sigma}:\mathbb{R}^n\times U\times\mathbb{R}^q\to\mathbb{R}^{n\times n} \) define the uncertain drift and diffusion coefficients, respectively, and $U\subset\mathbb{R}^m$ is a compact control constraint set.  The \( q \)-dimensional random variable \( {\bm \xi}:\Omega\to\mathbb{R}^q \), with randomness $\omega \in \Omega$, represents the system's uncertainties, and \( \bar{\bm \xi} \) denotes its expected value.

Based on the  \eqref{eq:SDE} form and \citep{lew2023risk}, an OCP is defined as
\begin{subequations}
    \begin{align}
    {\textsc{OCP}}:\ \inf_{{\bm u}\in\mathcal{U}}&\ \mathbb{E}\left[\int^{t_f}_0 \ell({\bm x}_{\bm u}(t),{\bm u}(t))dt + \varphi({\bm x}_{\bm u}(t_f))\right],\\
    {\rm s.t.}\quad &{\rm CVaR}_\alpha \left(\sup_{\tau\in[0,t_f]} C({\bm x}_{\bm u}(\tau),{\bm \xi}(\omega))\right)\leq 0,\\ \label{eq:cvar_ocp}
    & \mathbb{E}\left[H({\bm x}_{\bm u}(t_f))\right] = 0,\\ \label{eq:expect_ocp}
    & {\bm x}_{\bm u}\ {\rm satisfies (SDE)},
\end{align}
\label{eq:firstOCP}
\end{subequations}
where  $\ell:\mathbb{R}^n\times U \to\mathbb{R}$ is a running cost, $\varphi:\mathbb{R}^n\to\mathbb{R}$ is a terminal cost, $\alpha\in(0,1]$ is a risk parameter and $\mathcal{U}$ is the control input space. 
For a random variable $Z:\Omega\to\mathbb{R}$, ${\rm CVaR}_\alpha(Z(\omega)) = \inf_{z\in\mathbb{R}}\left(z+\frac{1}{\alpha}\mathbb{E}[{\max(Z-z,0)}]\right)$.
Let ${\bm x}_{\bm u}$ be the unique solution of the \eqref{eq:SDE}, for every ${\bm u} \in \mathcal{U}$. Finally, $C:\mathbb{R}^n\times\mathbb{R}^q\to\mathbb{R}$ and $H:\mathbb{R}^n\to\mathbb{R}$ are constraint functions such as collision avoidance or target reaching.

The formulation of the OCP is subject to the following assumptions: Lipschitz continuity and boundedness of the drift \(\bm{b}\) and diffusion \(\bm{\sigma}\) coefficients; Lipschitz continuity and boundedness of the running cost \(\ell\), terminal cost \(\varphi\), and constraint functions \(G\) and \(H\);  compactness of the control space \(\mathcal{U} \subset L^2([0,t_f],U)\); and square-integrability of the initial state \({\bm x}_0\) and uncertain parameters \({\bm \xi}\).  Furthermore, \({\bm x}_0\) and \({\bm \xi}\) are assumed to be $\mathcal{F}_0$-measurable, meaning their values are determined by the information available in the initial sigma-algebra \(\mathcal{F}_0\) at time 0.
These assumptions are equivalent to (A1)-(A4) in \cite{lew2023risk}.
The control objective can be considered achieved if this OCP is successfully solved. In \cite{lew2023risk}, the OCP is solved using SAA, evaluating expected costs and constraints by computing trajectories based on sampled parameters. However, large parameter uncertainties cause scattered trajectories, complicating the OCP's numerical resolution in practice.

\section{Risk-Aware TRAJECTORY PLANNING AND CONTROL}
\label{sec:Method}

\subsection{Feedback-Augmented RA-SAA}
\begin{figure*}
\centering
        \centering
        \includegraphics[width=1.7\columnwidth]{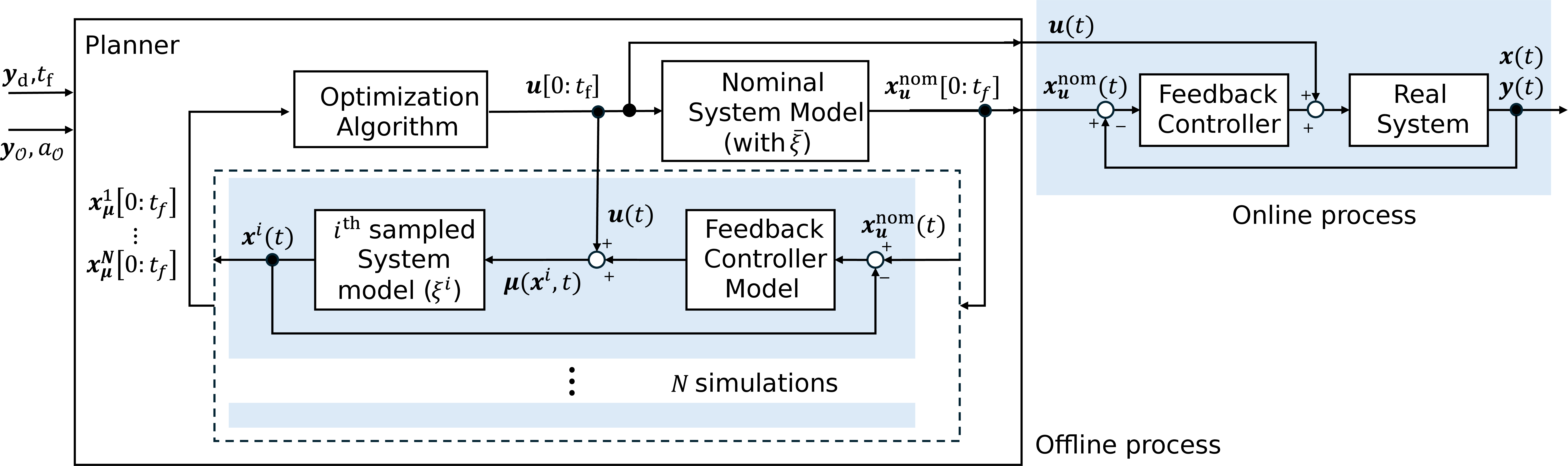}
    \caption{Overview of the RA-SAA+FB method. The bracket $[0:t_f]$ indicates a time series of the variable.}
    \label{fig:overview}
\end{figure*}

Our proposed method, denoted RA-SAA+FB, augments the RA-SAA framework with a feedback controller. The core idea is to center the uncertain trajectory realizations around a nominal, noise-free trajectory by adding a feedback term that compensates for deviations. This structure is illustrated in Figure \ref{fig:overview}.

We start by defining a nominal state ${\bm x}_{\bm u}^{\rm nom}(t)$ with respect to a control ${\bm u}\in \mathcal{U}$ and the expected value of the uncertain parameters distribution $\bar{\bm \xi}$
\begin{equation}
    d{\bm x}_{\bm{u}}^{\rm nom}(t) = {\bm b}({\bm x_{\bm{u}}}^{\rm nom}(t),{\bm u}(t), \bar{\bm \xi})dt\label{eq:SDE_nominal}, \quad \bm  {\bm x}_{\bm{u}}^{\rm nom}(0)  = \bm x_0.
\end{equation}
for $t \in [0,t_f]$ and an initial condition $\bm{x}_0$. We then introduce the control law ${\bm \mu}(\bm x^i, t)$\footnote{Regarding the notation, ${\bm u}$ is a pre-computed plan that defines the control law ${\bm \mu}(\bm x, t)$, rather than being an explicit argument.} for each sampled system model $i\in\{1,\dots,N\}$ with parameter ${\bm \xi}^i = {\bm \xi}(\omega^i)$, consisting of feedforward input ${\bm u}(t)$ and a correction feedback term scaled by gain matrix $\bm{K}$:
\begin{equation}
    {\bm \mu}(\bm x^i,t) = {\bm u}(t) + \bm{K}({\bm x}_{\bm{u}}^{\rm nom}(t)-{\bm x}^i(t))\label{eq:ex_linear_FB}.
\end{equation}
where $\bm x^i(t)$ is the state of the sampled model $i$ in the offline process at time $t$, or the measured state of the real system in the online process $\bm x(t)$, cf. Figure \ref{fig:overview}.  
The addition of this feedback controller is motivated by two key considerations: 1) to keep trajectory realizations more tightly centered around the nominal trajectory ${\bm x}_{\bm u}^{\rm nom}$ to make the constraint satisfaction easier in the OCP 2) allowing to solve the OCP with a lower frequency (as it requires more computation time) while running the feedback controller locally at a higher rate. 
The feedback gain $\bm{K}$ can be derived from standard robust control methods or heuristics.  High gains yield faster tracking and disturbance rejection but increase noise sensitivity and control effort. Low gains result in slower correction but often offer smoother control with lower effort and reduced risk of instability.
Note that the tracking error ${\bm x}_{\bm u}^{\rm nom}(t)-{\bm x}^i(t)$ does not necessarily have to vanish, but smaller tracking errors aid in finding valid solutions for the OCP, enabling a potential increase in $t_f$. For $\bm{K}=\bm{0}$, the method defaults to  RA-SAA.

Inserting \eqref{eq:ex_linear_FB} into the \eqref{eq:SDE}, the closed-loop trajectory of sample $i$ yields
\begin{equation}
\begin{split}
        d{\bm x}^i_{\bm{\mu}}(t) &= {\bm b}\left({\bm x}^i(t),{\bm \mu}(\bm x^i, t), {\bm \xi}^i\right)dt \\
        &\quad + \bm{\sigma}({\bm x}^i(t),{\bm \mu}(\bm x^i,t), {\bm \xi}^i)d{\bm W}_t.
\end{split}
\label{eq:expand_SDE}
\end{equation}
Our objective, is to find an optimal open-loop control plan ${\bm u}(t)$ by evaluating its cost over simulated trajectories $\bm{x}_{\bm \mu}^i$, generated from sampled model parameters, and under the influence of the control law ${\bm \mu}(\bm x^i, t)$ in \eqref{eq:ex_linear_FB}, later to be used in the online phase. In a nutshell, even though the optimization variable is $\bm u$, we are optimizing over closed-loop trajectories, given a fixed correction feedback gain $\bm{K}$. Provided $N$ independent and identically distributed samples $\omega^i \in \Omega$, we formulate a \emph{tractable deterministic approximation} of the (stochastic) OCP \eqref{eq:firstOCP} via Sample Average Approximation (SAA):
\begingroup
\makeatletter
\tagsleft@true
\begin{align}
\tag{${\textsc{SOCP+FB}}$}\label{eq:augSOCP}
\end{align}
\makeatother
\endgroup
\vspace{-0.4cm}
\begin{subequations}

\begin{align}
\inf_{{\bm u}\in\mathcal{U}} &\; \frac{1}{N}\sum_{i=1}^N \int_0^{t_f} \ell\bigl({\bm x}_{\bm \mu}^i(t),{\bm \mu}(\bm x^i, t)\bigr)\,dt + \varphi\bigl({\bm x}_{\bm \mu}^i(t_f)\bigr),\\
{\rm s.t.} \, \inf_{s\in\mathbb{R}} & \, s+ \frac{1}{\alpha N}\sum_{i=1}^N\max\Bigl(\sup_{\tau\in[0,t_f]}  C\bigl({\bm x}_{\bm \mu}^i(\tau),{\bm \xi^i}\bigr)-s,0\Bigr)\le 0 ,\label{eq:G_socp} \\
&\; -\delta_M \le \frac{1}{N}\sum_{i=1}^N H\bigl({\bm x}_{{\bm \mu}}^i(t_f)\bigr) \le \delta_M, \label{eq:H_socp} \\ 
 & \bm{\mu}, {\bm x}_{\bm \mu}^i \text{ satisfy \eqref{eq:ex_linear_FB} and \eqref{eq:expand_SDE} respectively}, \; \forall i \in [N].
\end{align}
\end{subequations}
where $\delta_M$ is a tolerance between the system's final position and the target, allowing small deviations while keeping the problem feasible \cite{lew2024sample}.

\subsection{Implementation}\label{sec:implementation}
We map the RA-SAA+FB method to our UUV-USV model described in Section \ref{sec:modeling}. While the time-discretized version of \eqref{eq:augSOCP} with an Euler-Maruyama discretization of \eqref{eq:SDE} is skipped in the interest of space, it follows directly from \citep{lew2023risk}.
The uncertain parameter ${\bm \xi}$ is set as
\begin{equation}
        {\bm \xi}(\omega) = \left(\begin{array}{c}
             (m(\omega),\bar{m}(\omega),c_\theta(\omega),c_r(\omega))^\top \\
             (\dots,x_{\mathcal{O}_k}(\omega),\dots)^\top\\
             (\dots,d_{\mathcal{O}_k}(\omega),\dots)^\top\\
             (\dots,a_{\mathcal{O}_k}(\omega),\dots)^\top
        \end{array}\right),
        \label{eq:xi}
    \end{equation}
and using the system expression \eqref{eq:input_affine},  the drift and diffusion terms of the \eqref{eq:SDE}  are given by

\begin{align*}
    &{\bm{b}} \left({\bm x}(t),{\bm u}(t), {\bm \xi}\right) = \bm{f}_{sys}({\bm x}(t),\bm \xi) + \bm{G}_{sys}({\bm x}(t),\bm \xi){\bm u}(t), \\
    & {\bm{\sigma}} \left({\bm x}(t),{\bm u}(t), {\bm \xi}\right) = \bm{D}
\end{align*}
where $\bm{D}$ is a matrix that determines the gain multiplying the Wiener process. Next, we define the cost functions ${\ell},{\varphi}$ and the constraint functions $ C, H$.
\begin{align}
    &\ell\bigl({\bm x}_{\bm \mu}^i(t),{\bm \mu}(\bm x^i, t)\bigr) = \|{\bm \mu}(\bm x^i, t)\|^2_R, \quad \varphi( {\bm x}^i_{\bm \mu}(t_f)) = 0 \label{eq:RA-SAA_cost_func}\\
    & C( {\bm x}_{\bm \mu}^i(t),{\bm \xi^i}) = 
    \max_k \left(-\left\|\left(\begin{array}{c}x^i(t)-x^i_{\mathcal{O}_k}\\d^i(t)-d^i_{\mathcal{O}_k}\end{array}\right)\right\| + a^i_{\mathcal{O}_k} \right), \label{eq:RA-SAA_collision_constraint}\\
    & H( {\bm x}^i_{\bm \mu}(t_f)) = \|{\bm y^i_{\bm \mu}}(t_f)-{\bm y}_{d}\|^2. \label{eq:RA-SAA_terminate_constraint}
\end{align}
with the obstacle positions $(x_{\mathcal{O}},d_{\mathcal{O}})$ and size $a_{\mathcal{O}}$, $k$ the index of each obstacle and $\bm{y}=(x, d, X)^\intercal$ as defined in \eqref{eq:input_affine}. The cost function defined in Equation \eqref{eq:RA-SAA_cost_func} minimizes the energy associated with the UUV-USV control inputs with penalization weight matrix $R$.  Equation \eqref{eq:RA-SAA_collision_constraint} defines $C$ as the signed distance to the nearest obstacle, which is negative (and therefore safe) whenever the UUV is outside all obstacles, and inserts it into the CVaR constraint \eqref{eq:G_socp} to enforce collision avoidance. Equation \eqref{eq:RA-SAA_terminate_constraint} represents the distance to the target point at the terminal time, and the constraint in Equation \eqref{eq:H_socp} ensures that the target position is reached.
Finally, we apply a (smooth) saturation function $\texttt{sat}(\cdot)$  that limits the control input within $U$, such that $\bm{\mu}$ reads
\begin{align}
    {\bm \mu}(\bm x^i,t) = \texttt{sat} \left({\bm u}(t) + \bm{K}({\bm x}_{\bm{u}}^{\rm nom}(t)-{\bm x}^i(t))\right),
\end{align}
The control law ${\bm \mu}$ is Lipschitz continuous, as the norm of $\bm{K}$ is finite, and bounded, due to the saturation function. Since the system and cost functions (${\bm{b}},{\bm{\sigma}}, \ell, \varphi, G, H$) from Section \ref{sec:modeling} also meet standard assumptions (A1)-(A2), the same asymptotic optimality results can be derived as in \citep{lew2023risk}, thus providing convergence guarantees to the optimal solution of \eqref{eq:augSOCP} in the limit $N \to \infty$ of the sample size.


\section{Simulation}
\label{sec:Sim}

In this section, we aim to validate our proposed RA-SAA+FB method through MATLAB simulation.
First, we consider an $8({\rm m})\times 8({\rm m})$ space with an obstacle.
We uniformly randomized the initial points ${\bm y}_0$, target points ${\bm y}_d$, as well as obstacle parameters $(x_{\mathcal{O}},d_{\mathcal{O}})$ and $a_{\mathcal{O}}$.
If the start or initial point is too close to the obstacle position or the water surface, these points are resampled. 
$N_{\rm location}$ denotes the number of such randomized scenarios.

We also randomized the weight of the UUV with litter $m$, its underwater weight $\bar{m}$ and the drag coefficient $c\ (= c_r = c_\theta).$
The randomization of these system parameters follows uniform distributions with expected values obtained from camera images or sensors, and a base range given by $\epsilon$ times the expected value, related to the relative error of the estimation. $N_{\rm model}$ denotes the number of such randomized models. The model parameters are listed in Table \ref{tab:model_parameters}. $(x_{\mathcal{O}},d_{\mathcal{O}})$ and $a_{\mathcal{O}}$ have 30\% and 10\% relative observation error, respectively.

In this simulation, we use a uniform distribution for the occurrence probability of uncertain parameters, reflecting limited certainty, as all deviations from the expected value are equally likely. Alternatively, a normal distribution could be used, where the probability density decreases with deviation, indicating greater knowledge or certainty about the expected value.


\begin{table*}
    \caption{Model Parameters in simulation}
    \label{tab:model_parameters}
    \centering
    \begin{tabular}{ccccl}
        Character & Value & Unit & Uncertainty &Description \\ \hline
        $m, \bar{m}, M, M_l$ & 120, 90, 1075, 30 & kg & $\pm120\epsilon, \pm90\epsilon, -, -$ & Masses: UUV (air), UUV (apparent), USV, winch \\
        $g$ & 9.8 & ${\rm m/s}^2$ & $-$ & Gravitational acceleration \\
         $c = c_r = c_\theta$ & 120 & ${\rm N}\cdot {\rm s^2/m^2}$ & $\pm120\epsilon$ & Drag parameter for the UUV\\
        $c_l, c_X$ & 300, 1000 & ${\rm N}\cdot {\rm s/m}$ & $ -, -$ & Viscosity (winch, USV) parameters \\
        $u_{max}$ & 400 & N & $-$ & Maximum thruster force \\
        $T_r, T_\theta, T_l, T_X$ & 0.1, 0.1, 0.5, 1.0 & s & $-$ & Time delays from control input to forces \\\hline
    \end{tabular}
\end{table*}
\begin{table}
    \caption{Control Parameters of RA-SAA and RA-SAA+FB in simulation}
    \label{tab:model_parameters_control}
    \centering
    \begin{tabular}{ccl}
        Character & Value &Description \\ \hline
        $R$ & $1/m^2\cdot I_{4\times 4}$ & input cost weight\\
        $\alpha$ & 0.02 & risk parameter\\
        $\delta_M$ & 0.3 & slack variable\\
        $\bm{K}_{\rm P}$ & $800\cdot I_{4\times 4}$ & proportional control gain\\
        $\bm{K}_{\rm D}$ & $80\cdot I_{4\times 4}$ & derivative control gain\\\hline
    \end{tabular}
\end{table}

\subsection{Baseline Approaches}
In $N_{\rm location}\times N_{\rm model}$ sets of scenario, we compared RA-SAA+FB to following three methods:
\begin{itemize}
    \item RA-SAA (\cite{lew2023risk}): offline planning and direct input application to the system.
    \item $A^*$+PID+CBF: generate a reference trajectory by the $A^*$ method, which only considers spatial relationships and does not take the system's dynamics into account. PID controller to follow a reference trajectory and CBF (\cite{ames2017control}) to guarantee safety w.r.t obstacles.
    \item MPPI (\cite{,williams2017model,williams2018information}): online evaluation of the predicted trajectory with sampled control inputs and rollout of the optimal control input to the system.
\end{itemize}
The methods discussed here plan and compensate for uncertainty either during the planning phase or via online feedback but differ in approach. The A* algorithm computes obstacle-avoiding paths by minimizing Euclidean distance, and when paired with PID and CBF, it blends offline planning with real-time adjustments.
The RA-SAA and Model Predictive Path Integral Control (MPPI) approaches easily handle nonlinear models, such as our UUV-USV model. Although both rely on stochastic trajectory sampling, RA-SAA operates in an open-loop manner and incorporates uncertainty in CVaR constraints entirely during offline planning, whereas MPPI uses importance sampling to determine optimal control inputs.

In RA-SAA and RA-SAA+FB methods, formulation of cost and constraint function follows (\ref{eq:RA-SAA_cost_func})-(\ref{eq:RA-SAA_terminate_constraint}). 
Adding to this, we used PD control for the feedback in our RA-SAA+FB method with gain matrices $\bm{K}_{\rm P}$ and $\bm{K}_{\rm D}$;
\begin{equation}
    \bm{K} = (\bm{K}_{\rm P}\ \bm{K}_{\rm D}\ O_{4\times 4}).
\end{equation}
Table \ref{tab:model_parameters} lists the value of these control parameters in our simulation. For the methods requiring solving an optimization problem, we use the \texttt{fmincon} solver in MATLAB, with initial solution guess ${\bm u} = (0,0,-\bar{m}g,0)^\top$. The time discretization step was set to 0.05 seconds.

Basic CBF and MPPI require deterministic models. In the implementations of the A*+PID+CBF method and the MPPI method, we used models based on equation \eqref{eq:input_affine} with the expected values of the parameters $\bar{\xi}$.

\subsection{Metrics}
To compare these methods, we define the following three metrics. Let $i\in\{1,\dots,N_{\rm location}\}$ and $\ j\in\{1,\dots,N_{\rm model}\}$ denote the indices of the location scenario and the sampled model respectively.
\begin{itemize}
\item Final position error: Evaluates the ability to reach exactly the target position, which for each method and each $i,j$ is defined as
    \begin{align*}
        \rho^{\rm final}_{ij} := \|{\bm y}(t_f)-{\bm y}_{d}\|.
    \end{align*}
    \item Collision rate: Evaluates the ability to avoid collisions. For each method, we define a collision flag $\rho^{\rm collision}_{ij}$ as follows
    \begin{align*}
        \rho^{\rm collision}_{ij} := \begin{cases}
        1 & \min_{t\in[0,t_f]} -\|{\bm y}(t)-{\bm y}_{\mathcal{O}}\|^2+\left(a_{\mathcal{O}}\right)^2 < 0\\
        0 & {\rm else}
        \end{cases}.
    \end{align*}
    \item Energy consumption: Evaluates the efficiency of the planned trajectory. For each method and for each $i,j$, we define the input energy $\rho^{\rm energy}_{ij}$ as follows
    \begin{align*}
        \rho^{\rm energy}_{ij} := \int_0^{t_f}\|{\bm f}(t)\|^2 dt.
    \end{align*}
\end{itemize}
with $\bm{f}$ is the force input defined above \eqref{eq:input_force_relation}. For each criterion, we calculate the mean value over $j$, i.e. $\rho_i = \frac{1}{N_{\rm model}}\sum_j \rho_{ij}$, 
then we compare the mean and variance value of $\rho_i$ for the different approaches.

We statistically compare the performance of RA-SAA+FB, given our defined metrics, to the other methods using a one-sided two-sample $t$-test with a significance level set to 0.05 (5\%). Statistically significant differences are marked with an asterisk (*) in Figure \ref{fig:result}.

\subsection{Results}
\begin{figure*}[ht!]
\centering
    \begin{minipage}{0.66\columnwidth}
        \centering
        \includegraphics[width=\columnwidth,trim=0.1in 0in 0.1in 0.1in, clip]{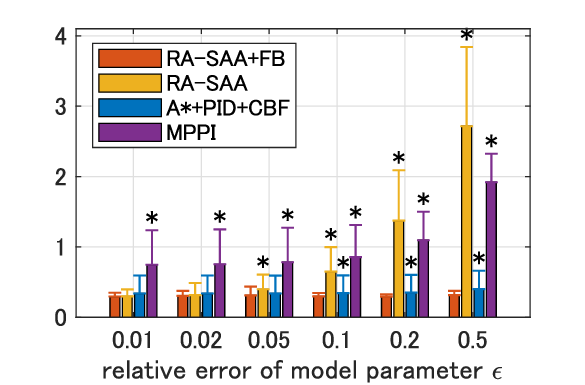}
        \subcaption{Final position error (${\rm m}$)}
        \label{fig:termination}
    \end{minipage}
    \begin{minipage}{0.66\columnwidth}
        \centering
        \includegraphics[width=\columnwidth,trim=0.1in 0in 0.1in 0.1in, clip]{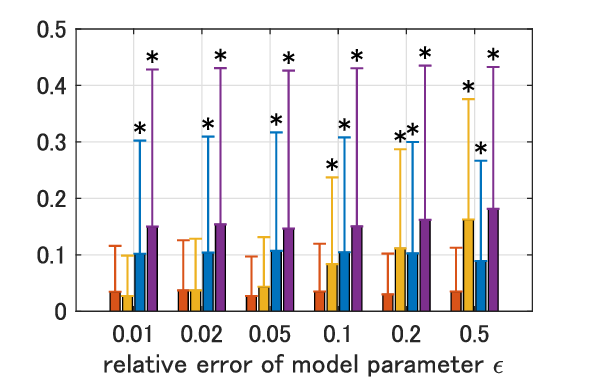}
        \subcaption{Collision rate}
        \label{fig:collision}
    \end{minipage}
    \begin{minipage}{0.66\columnwidth}
        \centering
        \includegraphics[width=\columnwidth,trim=0.1in 0in 0.1in 0.1in, clip]{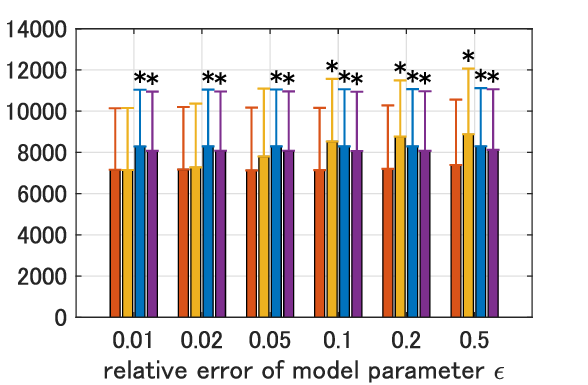}
        \subcaption{Energy consumption(${\rm N}^2\cdot {\rm s}$)}
        \label{fig:energy}
    \end{minipage}
    \centering
    \caption{Evaluation of metrics from the simulation results }
    \label{fig:result}
\end{figure*}
Figure \ref{fig:result} shows the simulation results with $N_{\rm location} = 100$, $N_{\rm model} = 20$. 
The horizontal axis represents the uncertainty range $\epsilon$, which determines the width of the distribution. 
The bar plots show the mean values of the metrics, with lines above each bar representing the standard deviations.

In Figure \ref{fig:termination}, we observe that the original RA-SAA and RA-SAA+FB performance is similar for small uncertainties. However, as uncertainty grows, the original RA-SAA's final position error increases due to infeasibility, while MPPI's performance also deteriorates. For $\epsilon = \{0.2, 0.5\}$, the final position error of RA-SAA+FB is slightly smaller than that of A*+PID+CBF.

For the collision rate shown in Figure \ref{fig:collision}, both RA-SAA and MPPI are sensitive to increased uncertainty.
RA-SAA+FB, on the other hand, is robust to parameter uncertainty. 
We also observe a slight increase in the collision rate for A*+PID+CBF.
At $\epsilon=0.5$, the collision rate of RA-SAA+FB is significantly lower than that of A*+PID+CBF.

Finally, we evaluate the control efficiency in Figure \ref{fig:energy}.
In all cases, RA-SAA+FB consumes the same or less input energy than other methods.

\subsection{Discussion}

\begin{figure}[!t]
\centering 
\begin{tabular}{cc}
    \begin{minipage}{0.45\hsize}
        \centering
        \includegraphics[width=2.1in]{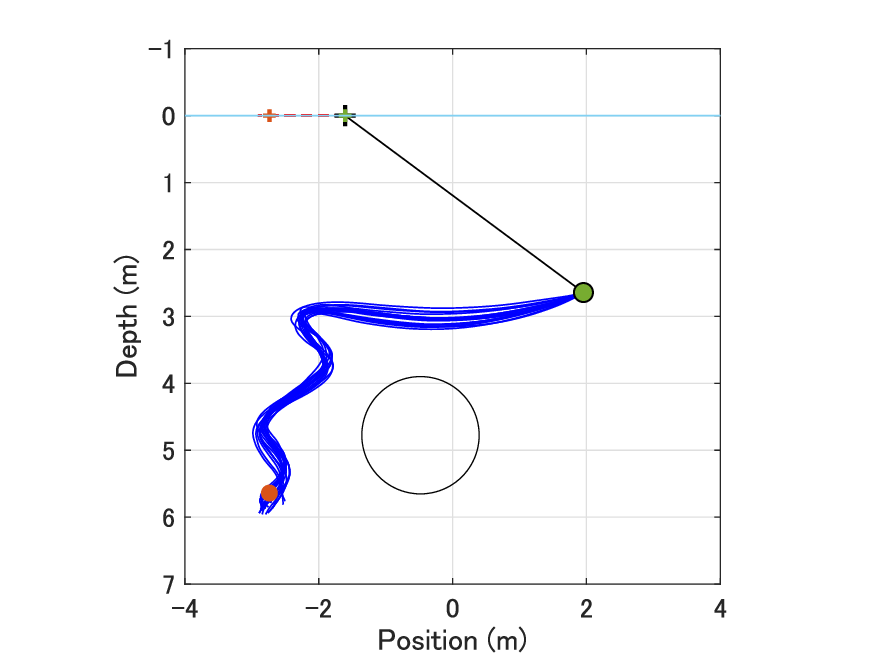}
        \subcaption{RA-SAA+FB}
        \label{subfig:ra_saa_fb} 
    \end{minipage}& 
    \begin{minipage}{0.45\hsize}
        \centering
        \includegraphics[width=2.1in]{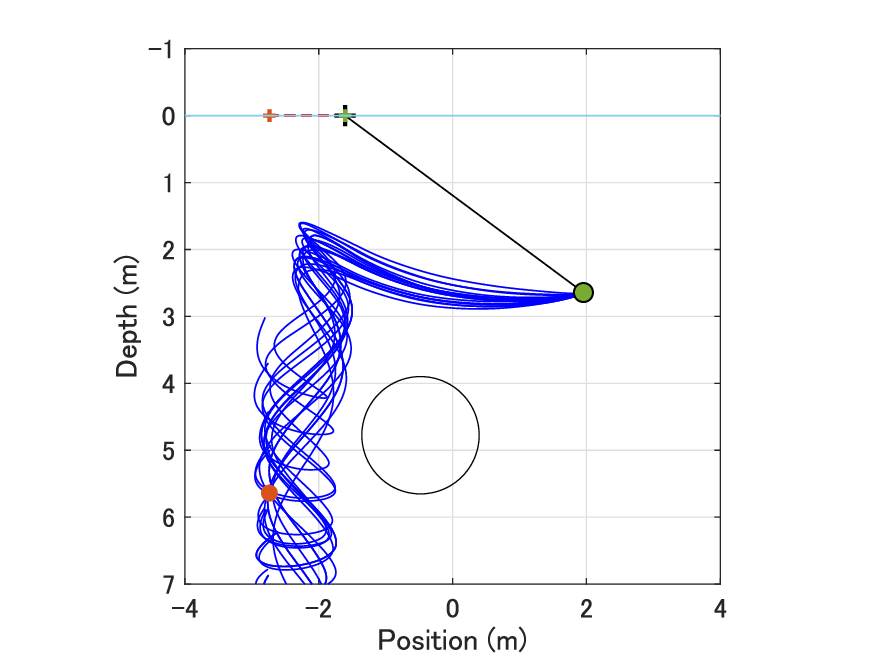}
        \subcaption{RA-SAA}
        \label{subfig:ra_saa} 
    \end{minipage}\\
    \begin{minipage}{0.45\hsize}
        \centering
        \includegraphics[width=2.1in]{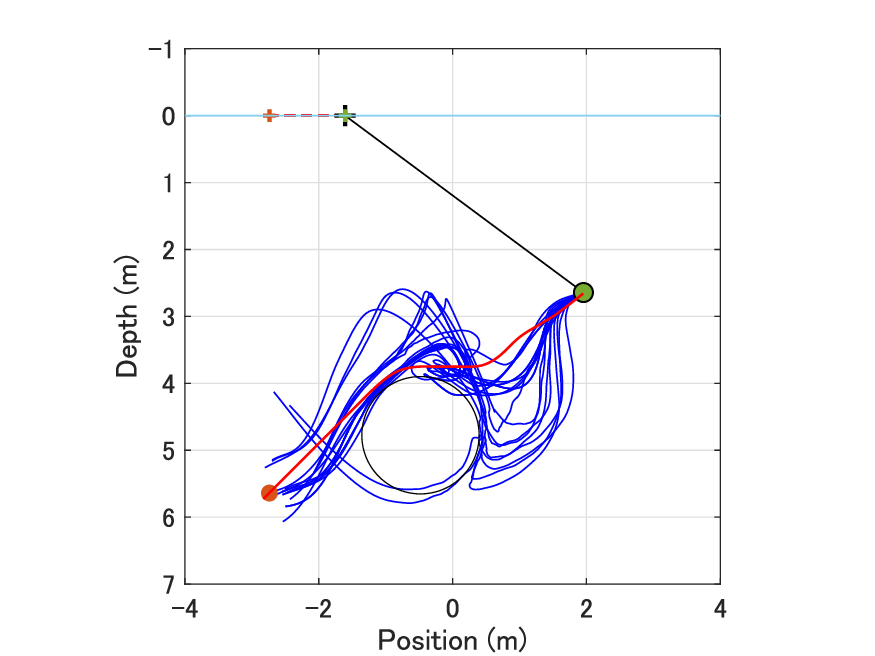}
        \subcaption{A*+PID+CBF}
        \label{subfig:astar_cbf} 
    \end{minipage}&
    \begin{minipage}{0.45\hsize}
        \centering
        \includegraphics[width=2.1in]{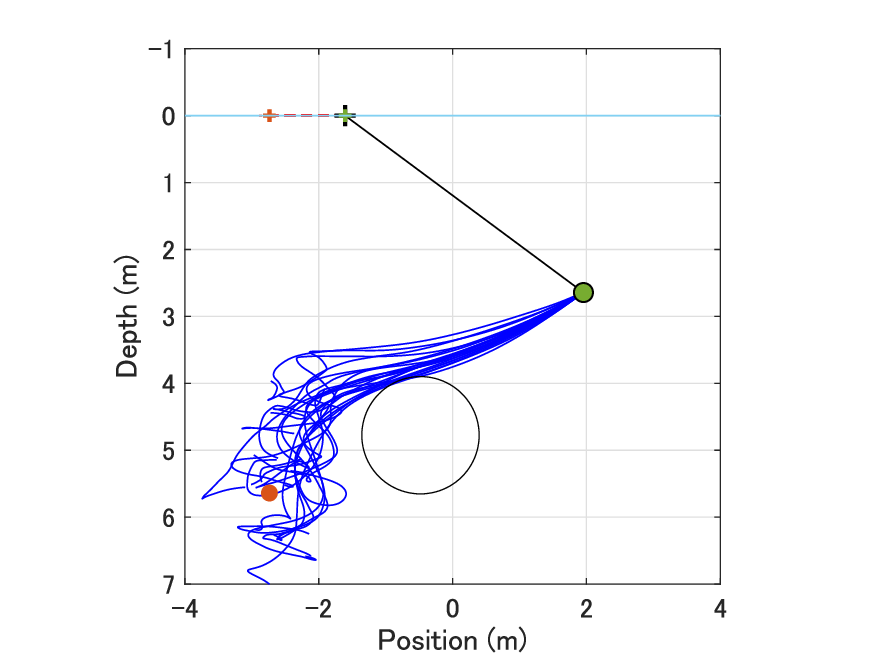}
        \subcaption{MPPI}
        \label{subfig:mppi} 
    \end{minipage}
\end{tabular}
\caption{Comparison of trajectories for $i=81$.}
\label{fig:trajectory}
\end{figure}

Figure \ref{fig:trajectory} illustrates the apparent performance differences between the methods reported in Fig. \ref{fig:result}. Green and orange points mark the UUV's initial and target positions, respectively; crosses denote the USV's positions, and the blue curves show the UUV's trajectory for each sampled model. Our proposed RA-SAA+FB method (Fig.~\ref{subfig:ra_saa_fb}) demonstrates superior performance, with trajectories converging to the target and compensating for litter weight and drag uncertainties across different sampled scenarios. The paths exhibit a natural pendulum-like motion, indicating that the optimization correctly aligns with the system's inherent dynamics. In contrast, the original RA-SAA method (Fig.~\ref{subfig:ra_saa}), which lacks the feedback controller, produces scattered trajectories. This comparison highlights the feedback term's critical role in centering the trajectories around the nominal model.

The baseline methods struggle with the system's dynamics and uncertainties. The A*+PID+CBF approach (Fig.~\ref{subfig:astar_cbf}) follows a kinematically optimal but dynamically unnatural straight-line path. This requires large control inputs to counteract gravity and, combined with a Control Barrier Function that does not consider system uncertainty, leads to frequent collisions. While the model-based MPPI controller (Fig.~\ref{subfig:mppi}) generates more natural-looking paths, its trajectories are still dispersed, and some result in collisions. This is because, unlike our method, its online planning does not account for uncertainty in obstacle positions, demonstrating the value of our risk-aware formulation.

\section{Conclusion}
In this paper, we proposed extending a risk-aware optimal control problem with CVaR-based constraints to solve trajectory optimization, integrating a local feedback controller to improve performance. We applied this approach to a novel UUV-USV litter collection system, enabling the system to achieve objectives such as reaching a target location and avoiding obstacles, even in the presence of significant parameter uncertainty. Using a Sample Average Approximation, we formulated a computationally tractable problem and applied it to a simulated planar model of the system. Results demonstrated that our approach outperforms standard CBF and MPPI control in collision avoidance and energy efficiency while improving on infeasibility issues in the original open-loop method.
Future work will extend the simplified planar model to three dimensions and incorporate environmental disturbances such as currents and waves before proceeding to real-world experimental validation.

\bibliography{references}
\end{document}